\documentclass[sigconf, 10pt]{acmart}

\renewcommand\footnotetextcopyrightpermission[1]{}

\AtBeginDocument{%
  }

\usepackage{amsmath}
\usepackage{algorithmic}
\usepackage{algorithm}
\usepackage{graphicx}
\usepackage{pifont}
\usepackage{textcomp}
\usepackage{xcolor}
\usepackage{booktabs} 
\usepackage{soul}
\graphicspath{{figs/}}
\usepackage{braket}
\usepackage{etoolbox}
\usepackage{fancyhdr}
\usepackage{multirow}
\usepackage{makecell}

\begin{document}

\fancyhead{}
\title{Inverse-Transpilation: Reverse-Engineering Quantum Compiler Optimization Passes from Circuit Snapshots}

\author{Satwik Kundu}
\affiliation{%
  \institution{Pennsylvania State University}
  \city{University Park}
  \state{PA}
  \country{USA}
  \postcode{16801}}
\email{satwik@psu.edu}

\author{Swaroop Ghosh}
\affiliation{%
  \institution{Pennsylvania State University}
  \city{University Park}
  \state{PA}
  \country{USA}
  \postcode{16801}}
\email{szg212@psu.edu}


\begin{abstract} 
Circuit compilation, a crucial process for adapting quantum algorithms to hardware constraints, often operates as a ``black box,'' with limited visibility into the optimization techniques used by proprietary systems or advanced open-source frameworks. Due to fundamental differences in qubit technologies, efficient compiler design is an expensive process, further exposing these systems to various security threats. In this work, we take a first step toward evaluating one such challenge affecting compiler confidentiality, specifically, reverse-engineering compilation methodologies. We propose a simple ML-based framework to infer underlying optimization techniques by leveraging structural differences observed between original and compiled circuits. The motivation is twofold: (1) enhancing transparency in circuit optimization for improved cross-platform debugging and performance tuning, and (2) identifying potential intellectual property (IP)-protected optimizations employed by commercial systems. Our extensive evaluation across thousands of quantum circuits shows that a neural network performs the best in detecting optimization passes, with individual pass F1-scores reaching as high as 0.96. Thus, our initial study demonstrates the viability of this threat to compiler confidentiality and underscores the need for active research in this area.
\end{abstract}

%
%



\maketitle

\section{Introduction}
The field of quantum computing is growing at an incredible pace, with several companies \cite{computing2023quantum, gambetta2023hardware} competing to design quantum hardware that can outperform classical computers on practically useful tasks. Recently, Google unveiled its 105-qubit superconducting chip, Willow \cite{acharya2024quantum}, and showed that it could complete random-circuit-sampling benchmark in about five minutes which is estimated to take Frontier, one of the world’s fastest supercomputers, roughly 10 septillion ($10^{25}$) years. Microsoft likewise announced its first quantum computer \cite{microsoft2025interferometric}, built from a novel ``topoconductor” based on Majorana particles, claiming that this platform is naturally more robust to noise and can scale to millions of qubits. As quantum hardware diversifies, the landscape of quantum compilers \cite{li2019tackling, chong2017programming} is also evolving, with each platform developing specialized tools to meet its technological demands. Because different devices exhibit distinct basis-gate sets, qubit-connectivity graphs, and noise profiles, designing efficient technology-specific quantum compilers is essential.


Quantum compilers play a crucial role in bridging high-level quantum algorithms and the physical hardware, and different qubit technologies, such as superconducting qubits, trapped ions, and photonic qubits which often require alternate compilation strategies and dedicated compilers. The compilation process involves translating abstract quantum circuits into hardware-executable instructions by performing a sequence of passes like, circuit optimization to reduce depth and gate count, qubit mapping and routing to satisfy connectivity constraints, gate decomposition into the device's native gate-set, etc. For example, IBM’s Qiskit compiler \cite{qiskit-transpiler} is tailored for superconducting qubits, optimizing circuits for IBM Quantum devices’ native gates and connectivity. In contrast, Quantinuum’s TKET \cite{sivarajah2020t} is designed to support a wide range of hardware backends, including trapped-ion systems, and offers advanced routing and optimization for those architectures. Xanadu’s Strawberry Fields \cite{killoran2019strawberry} compiler targets photonic quantum computers, focusing on continuous-variable operations and photonic-specific optimizations. Each compiler’s uniqueness lies in its ability to adapt quantum circuits to the strengths and limitations of the underlying qubit technology, ensuring efficient execution and maximizing fidelity on the chosen hardware. As a result, designing the optimal compilation strategy for any given qubit technology \cite{kreppel2023quantum, zou2024lightsabre} remains a non-trivial problem.

\begin{figure}[!t]
        \centering 
        \includegraphics[width=\linewidth]{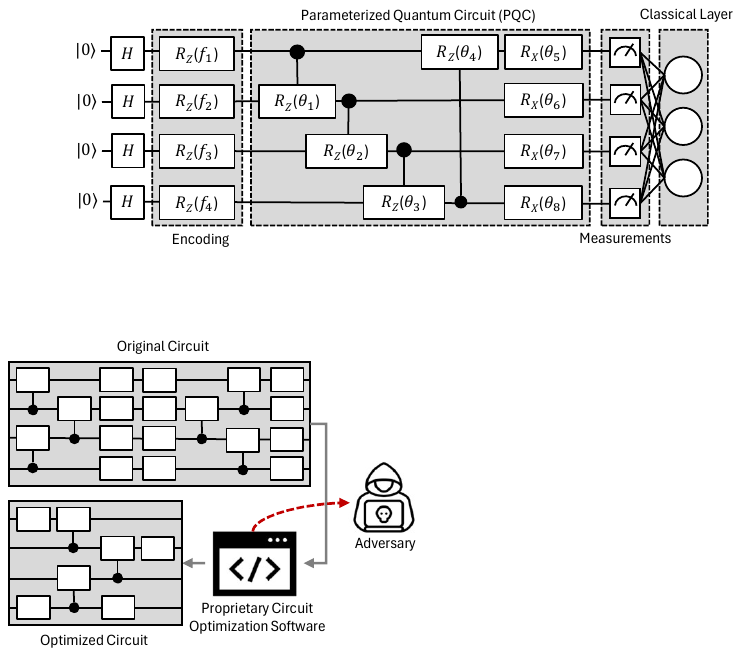}
        \vspace{-6mm}
        \caption{High-level overview of the threat model demonstrating a scenario where a stealthy adversary can infer details about black-box circuit-optimization software from the original and optimized circuits.}
        \label{threat_model}
        \vspace{-4mm}
\end{figure}

However, the rapidly evolving field of quantum hardware brings not only unique computational capabilities but also a variety of new security challenges \cite{kundu2024adversarial, tan2025qubithammer}. For instance, \cite{xu2024security} identifies multiple vulnerabilities at the gate-to-pulse interface of quantum circuits and proposes corresponding defense frameworks. Quantum compilers, a key part of the circuit-execution pipeline, have also been examined in several security-focused studies. The authors of \cite{saki2021split} introduced split compilation to protect circuit-encoded intellectual property from untrusted third-party compilers, while other work has reverse-engineered the original circuit \cite{ghosh2024quantum} and even the backend coupling map \cite{roy2024forensics} from the transpiled output. However, one critical threat which remains unaddressed is the risk to the intellectual property of the compilers themselves. Specifically, can an external adversary extract information about a compiler’s optimization techniques simply by comparing the original and transpiled circuits (Fig. \ref{threat_model})? If so, the compiler’s confidentiality would be seriously compromised.

In this study, we design a proof-of-concept framework that leverages machine-learning models to identify the optimization techniques employed by a black-box compiler solely from the original and optimized circuits. Our main contributions are,  
\begin{itemize}
    \item Introducing a previously unexplored security vulnerability of quantum compilers, namely, the extraction of gate-level optimization techniques.  
    \item Extracting intuitive but informative features from original and compiled circuits and training a model for multi-label classification to predict the possible strategies used by the compiler.  
    \item Evaluating multiple ML models on quantum circuits of varying width and depth, demonstrating their effectiveness in identifying the compiler’s optimization techniques with measurable confidence. 
\end{itemize}  
\textit{To the best of our knowledge, this is the first work to infer a quantum compiler’s optimization strategies from the abstract and optimized circuits.}

\section{Background } \label{background}

\subsection{Basics of Quantum Computing}  
A quantum bit, or qubit, is the fundamental building block of quantum computers and is typically driven by microwave pulses for superconducting qubits or laser pulses for trapped-ion qubits. Unlike a classical bit, a qubit can be in a superposition state, representing a combination of $\ket{0}$ and $\ket{1}$ simultaneously. Mathematically, a qubit state is represented by a two-dimensional column vector $\left[\begin{smallmatrix}\alpha \\ \beta\end{smallmatrix}\right]$, where $|\alpha|^2$ and $|\beta|^2$ represent the probabilities of the qubit being in state `0' and `1', respectively. Quantum gates are operations that manipulate qubit states and are represented by unitary matrices (e.g., Pauli-Z gate: $\left[\begin{smallmatrix}1 & 0 \\ 0 & -1\end{smallmatrix}\right]$). There are two main types of quantum gates: single-qubit gates (e.g., H, X) and two-qubit gates (e.g., CNOT, CRY). More complex gates, such as the three-qubit Toffoli gate, are typically decomposed into sequences of one- and two-qubit gates during compilation. Qubits are measured in a chosen basis to determine the final state of a quantum program. In physical quantum computers, measurements are usually restricted to the computational basis, such as the Z-basis used in IBM systems. Due to high error rates, single measurement outputs are generally inaccurate. Therefore, quantum circuits are measured multiple times ($n$ shots), and the most frequent outcomes are considered the final results.

\subsection{Quantum Circuit Compilation} \label{compilation}
Quantum‐circuit compilation \cite{li2019tackling, quetschlich2025mqt, ruiz2025quantum} transforms an algorithm level circuit, expressed in an idealized gate set, into an implementation that can run efficiently on a specific quantum processor. This transformation is necessary because real hardware supports only a few native gates, offers limited qubit connectivity, and exhibits various noise and timing characteristics. A compiler therefore tackles four core tasks: (i) gate decomposition, rewriting any non-native gate into an equivalent sequence of supported one- and two-qubit gates; (ii) qubit layout, assigning logical qubits to physical qubits so as to minimize a cost that mixes connectivity distance, gate fidelities, and coherence times; (iii) routing, which introduces SWAP operations when required interactions involve unconnected qubits; and (iv) scheduling and optimization, which adjust gate start times, cancel redundant operations, and merge commuting sub-circuits to shorten depth and reduce cumulative error.

Currently, most libraries implement this workflow as a sequence of passes. In Qiskit \cite{qiskit-transpiler}, for example, layout begins with a graph-isomorphism search (VF2Layout) to find a connectivity-preserving assignment. When that fails, a heuristic pass (SabreLayout/SabreSwap) iteratively swaps qubits while simulating circuit execution to minimize the number of added SWAP gates. Translation to the native basis relies on an equivalence-graph search that applies Dijkstra’s algorithm to identify the lowest-cost decomposition of each unsupported gate, and repeated peephole passes eliminate unnecessary single- and two-qubit gate pairs introduced during earlier stages. All passes consult a target object that encodes per-gate error rates and durations, enabling noise-aware cost models. The output is a scheduled, hardware-compatible circuit that typically exhibits a substantially lower depth and lower aggregate error than a naive mapping, thereby extending the problem sizes and algorithmic depths that near-term quantum computers can handle.

\subsection{Related Works} 
Security of quantum systems has recently garnered a lot of attention \cite{rasmussen2024time, choudhury2024crosstalk, chen2024nisq}, with multiple studies examining different stages of the quantum-as-a-service (QaaS) pipeline. In the context of compilation, prior work \cite{ghosh2024quantum} has reverse-engineered quantum-machine-learning circuits to reconstruct the original circuits from their transpiled forms. Other studies \cite{roy2024forensics} have inferred a backend’s coupling map from the transpiled output. Researchers have also proposed defense techniques \cite{saki2021split} such as circuit splitting to protect proprietary algorithms from untrusted compilers. However, no existing work has focused on extracting information about the optimization techniques employed by a compiler.



\section{Adversarial Framework}

\subsection{Threat Model}

We consider a vendor that has developed a state-of-the-art quantum-circuit optimization algorithm and provides paid access to its proprietary optimization software. We examine a threat model in which an unknown adversary uses this software to obtain gate-optimized versions of original circuits. The main concern is confidentiality: by comparing the structures of the input and the optimized output, could the adversary infer details about the provider’s optimization methods? If so, the confidentiality of the framework would be severely compromised.

The adversary’s primary goal is to identify, with some level of confidence, the circuit-optimization rules or techniques employed by the proprietary algorithm. The adversary may be motivated by various factors, such as (1) gaining a competitive market advantage, (2) revealing insights that facilitate intellectual-property (IP) theft, (3) selling the stolen IP for profit, or (4) using the extracted rules to build a competing circuit optimizer and offer a similar service. Ultimately, the adversary aims to retrieve algorithmic information about the provider’s proprietary methods.

\begin{figure}[!t]
        \vspace{-4mm}
        \centering 
        \includegraphics[width=\linewidth]{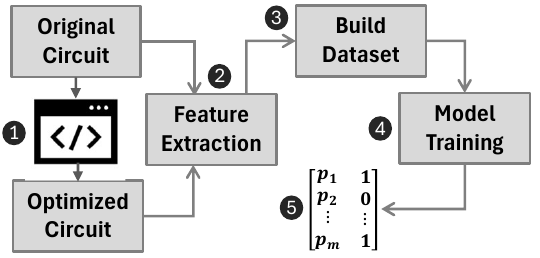}
        \vspace{-5mm}
        \caption{Pipeline Overview: (1) Circuit pairs are generated based on a few preselected optimization passes; (2) High-level features are extracted from both circuits; (3) A dataset is built using these extracted features, where the input consists of features from the original and optimized circuits, and the label is a one-hot encoding of the optimization techniques (OT) applied; (4) A machine-learning model, such as a random forest, is trained on this dataset to perform multi-label classification (one-vs-all); (5) The model predicts the optimization passes ($p_i$) used by the compiler.}
        \label{fig:framework_pipeline}
        \vspace{-4mm}
\end{figure}

\subsection{Adversary Knowledge}

We assume that the adversary already knows several well-established optimization techniques and seeks to determine whether the proprietary algorithm applies any of them by comparing structural differences between the original and optimized circuits. The provider’s software is a black box: users (including the adversary) have no visibility into its internal workings. Access is granted via an API through which a user submits an original circuit and receives a gate-optimized version in return. Consequently, the adversary trains an ML-based model to infer details about the proprietary algorithm using only the input (original) and output (optimized) quantum circuits.

\section{Proposed Methodology} 

There are multiple stages in the quantum circuit compilation process, including gate decomposition, qubit mapping, and scheduling (Sec. \ref{compilation}). In this work, we primarily focus on the gate decomposition and optimization stages. Specifically, given a novel circuit optimization algorithm, we develop an ML-based technique that predicts the underlying optimization methods employed by the algorithm. Fig. \ref{fig:framework_pipeline} provides a high-level overview of our pipeline. 


\subsection{Optimization Pass-selection}
Even though we narrowed down to a single stage of the transpilation process, there are a variety of optimization techniques available and thus it is practically infeasible to train a model on all optimization rules. This is because, simply including every optimization pass would create hundreds of sparsely observed labels, rendering prediction task hopeless. Instead, we hand-picked few passes (like \texttt{GateDecomposition}, \texttt{CommutativeCancellation}), which cover the spectrum of optimization a developer is most likely to enable in production and call these \texttt{Baseline} passes. The idea here is that each pass injects a distinct structural fingerprint into the circuit (e.g. \texttt{InverseCancellation} predominantly reduces circuit depth) in atleast one of three dimensions:  gate-count, circuit depth, two-qubit count. By forcing the classifier to learn important orthogonal characteristics, the model can more accurately detect underlying optimization techniques.

However, real-world compilers, or in this scenario the proprietary circuit optimization software could have inserted passes which are novel and unknown to the user (or adversary in this scenario). To better mimic that black-box uncertainty we keep few popular but unlabeled passes in a separate bucket which we call~\texttt{Miscellaneous}. With a tunable probability, we randomly insert few of these into the pipeline. Whenever that happens the sample receives a single tag~\texttt{Miscellaneous}. This strategy trains the model to basically learn cases where some other optimization technique might have been used outside of the models knowledge, very much like our attack scenario where the proprietary software might have better performance due to the presence of few novel unknown optimization techniques.

\subsection{Circuit Generation}

With the optimization passes fixed, we now need to generate the circuits. Current quantum computers workload mostly falls into \textit{variational} or \textit{randomized} category, so instead of relying on a small set of algorithmic benchmarks, we generate thousands of \textit{random circuits} using Qiskit's functionality, each with $[q_{\min},q_{\max}]$ qubits and circuit depth $d$. The primary reason for choosing random circuits is because they cover the state space much more uniformly than hand-picked algorithms, preventing the model from over-fitting and prior works \cite{haferkamp2022random} have shown that such circuits already approximate unitary $t$-design, which means they expose the optimizer to the same tangled interaction that is present in near-term experiments.

Each circuit is duplicated i.e. one copy remains untouched (\textit{original}), the other is fed through a \texttt{PassManager} that executes a randomly selected, non-empty subset of baseline passes and, potentially, an additional set of miscellaneous passes. The optimization choices made in this step become the ground-truth one-hot label vector that the classifier is trained on.

\subsection{Features, Dataset and Training}
The feature selection step is crucial in our methodology. Here, we take a simple yet practical approach to extract important features from the circuits. Instead of hand-coding complex algebraic signatures, we use a collection of straightforward, human-interpretable statistics. Specifically, we consider global features such as depth, total gate count, and circuit width, as well as individual gate counts and aggregate counts (e.g., single-qubit and two-qubit gate counts). We chose these features because they reflect how a human expert might differentiate two circuits side-by-side. By presenting the original and optimized versions of the circuit separately as inputs during training, we allow the model to discover patterns independently. 

After the feature selection step, we obtain the final dataset in the form $(X, Y)$, where $X = (\texttt{original}, \texttt{optimized})$ and $Y$ is a one-hot representation of the baseline and miscellaneous optimization techniques (Algo. \ref{alg:dataset_builder}). We use this dataset to train machine learning models on a multi-label prediction task, where the model estimates the optimization technique which might have been used in the compilation process.


\begin{algorithm}[!t]
\caption{Synthetic Dataset Builder}
\label{alg:dataset_builder}
\begin{algorithmic}[1]
  \REQUIRE Sample count $N$; qubit range $[q_{\min},q_{\max}]$; circuit depth $d$; baseline pass-set $\mathcal{P}_b$; miscellaneous pass-set $\mathcal{P}_m$; misc probability $p$
  \ENSURE Feature matrix $\mathbf{X}$ and one-hot label matrix $\mathbf{Y}$

  \STATE $\mathbf{X},\mathbf{Y} \gets \varnothing$ \COMMENT{Initialize empty datasets}

  \FOR{$i = 1 \;\textbf{to}\; N$}
     \STATE $n_q \sim \mathcal{U}\bigl(q_{\min}, q_{\max}\bigr)$
     \STATE $\mathcal{C}_{\text{orig}} \gets \textsc{RandomCircuit}(n_q, d)$
     \STATE $\mathcal{S}_b \gets \textsc{RandomNonEmptySubset}(\mathcal{P}_b)$
     \STATE $\mathcal{S}_m \gets
        \begin{cases}
           \textsc{RandomSubset}(\mathcal{P}_m) & \text{with prob. } p \\
           \varnothing                           & \text{otherwise}
        \end{cases}$
     \STATE $\mathcal{C}_{\text{opt}} \gets
            \textsc{PassManager}\bigl(\mathcal{S}_b \cup \mathcal{S}_m\bigr)
            \bigl(\mathcal{C}_{\text{orig}}\bigr)$
     \STATE $x \gets \textsc{Features} \bigl(\mathcal{C}_{\text{orig}},
                                  \mathcal{C}_{\text{opt}}\bigr)$
     \STATE $y \gets \textsc{OneHot}(\mathcal{S}_b)$
            \COMMENT{Set $y_{\text{misc}}\!=\!1$ if $\mathcal{S}_m\!\neq\!\varnothing$}
     \STATE Append $x$ to $\mathbf{X}$; append $y$ to $\mathbf{Y}$
  \ENDFOR

  \RETURN $\mathbf{X}, \mathbf{Y}$
\end{algorithmic}
\end{algorithm}
\vspace{-2mm}

\section{Evaluation}
\subsection{Experimental Setup} \label{setup}
\noindent \textbf{Optimization Passes (labels):} Here, we specifically focused on 8 total optimization techniques provided by Qiskit, 6 of which are used as baseline optimization techniques and 2 techniques fall under the miscellaneous bucket.
The baseline techniques used are:
\begin{itemize}
    \item \texttt{Optimize1qGatesDecomposition}
    \item \texttt{InverseCancellation}
    \item \texttt{CommutationAnalysis} + \texttt{CommutativeCancellation} 
    \item \texttt{RemoveIdentityEquivalent}
    \item \texttt{ConsolidateBlocks}
    \item \texttt{TemplateOptimization}
\end{itemize}
The miscellaneous rules used are:
\begin{itemize}
    \item \texttt{Split2QUnitaries}
    \item \texttt{OptimizeCliffords}
\end{itemize}
We specifically chose these passes since they collectively span most major optimization techniques, such as local gate fusion, inverse/commutation cancellation, and block consolidation. Furthermore, these passes leave distinct, non-overlapping footprints (e.g., depth change, 1-qubit vs. 2-qubit gate counts), which can help the model better learn distinguishing characteristics of each technique. We used Qiskit's \texttt{random\_circuit()} functionality to generate circuits for our dataset based on the chosen optimization passes. Note that we randomly select baseline passes and include a \texttt{Miscellaneous} pass with probability 0.25 (since it comprises 2 techniques out of 8 total).

\begin{table}[!b]
    \vspace{-2mm}
    \centering
    \caption{Individual optimization pass performance on different evaluation metrics.}
    \label{tab:pass-performance}
    \vspace{-2mm}
    \begin{tabular}{cccc}
    \cmidrule(lr){1-4}
    Optimization Pass    & Precision & Recall & F1-Score  \\
    \cmidrule(lr){1-4}
    \texttt{Optimize1qGates}             & 0.80  & 0.75          & 0.77            \\
    \texttt{InverseCancellation}      & 0.55  & 0.46          & 0.49         \\
    \texttt{CommutationCancel}       & 0.50  & 0.49        & 0.49         \\
    \texttt{RemoveIdentity}       & 0.51  & 0.50          & 0.51         \\
    \texttt{ConsolidateBlocks}      & 0.49  & 0.48          & 0.49         \\
    \texttt{TemplateOptimization}      & 0.99  & 0.93          & 0.96         \\
    \texttt{Miscellaneous}      & 0.25  & 0.07          & 0.10         \\
    \bottomrule
    \end{tabular}
    \vspace{-2mm}
\end{table}

\noindent \textbf{Feature Selection:} For each circuit, we extract a compact, human-interpretable fingerprint comprising (i) five global descriptors: depth, total gate count, register width, qubit count, and circuit size, (ii) a histogram of the fourteen most frequent one-qubit and two-qubit gate types \{$u_1$, $u_2$, $u_3$, $R_z$, $X$, $H$, $S$, $S^{\dagger}$, $T$, $T^{\dagger}$, $CX$, $CZ$, SWAP, $RZZ$\}, and (iii) aggregate sums and ratios of one-qubit and two-qubit gates.

\noindent \textbf{Training Details:} For circuit generation, we predominantly used circuits with qubit counts in [4, 12] and a depth of 50. We evaluated the models on 10000 circuit samples, splitting them into training and test sets using an 80:20 ratio. We use three different evaluation metrics: precision, recall, and F1-score to measure performance on individual optimization passes; and Hamming score, average-F1, and micro-F1 to assess overall model performance. All training was performed on an Intel Core i7-12700H with 40GB of RAM.

\subsection{Results and Analysis} \label{sec:results}

\begin{table}[!b]
    \vspace{-2mm}
    \centering
    \caption{Optimization technique detection performance of different ML models evaluated using various metrics. Neural network outperforms the other models achieving the highest F1 and Hamming score.}
    \label{tab:benchmarks}
    \vspace{-2mm}
    \begin{tabular}{cccc}
    \cmidrule(lr){1-4}
    Model    & Hamming & Avg-F1 & Micro-F1  \\
    \cmidrule(lr){1-4}
    Neural Network                     & 0.682  & 0.594          & 0.624            \\
    Logistic Regression       & 0.662  & 0.573          & 0.609         \\
    Gradient Boosting       & 0.660  & 0.561        & 0.601         \\
    Random Forest           & 0.647  & 0.556          & 0.592         \\
    kNN (k = 5)                    & 0.616  & 0.528          & 0.568         \\
    \bottomrule
    \end{tabular}
    \vspace{-2mm}
\end{table}

\noindent \textbf{Pass-specific Performance:} Table \ref{tab:pass-performance} demonstrates the performance of individual passes using various evaluation metrics when training a random forest (with 300 decision trees). Here, precision tracks false positives and recall tracks false negatives, so a precision or recall of 1.0 indicates zero false positives or false negatives, respectively. From the table, our model is effective in detecting \texttt{Optimize1qGates} (F1-score 0.77) and \texttt{TemplateOptimization} (F1-score 0.96), indicating that our feature set captures reductions in single-qubit chains and template patterns with high confidence. In contrast, the F1-scores for the other baseline optimization passes are around 0.5, suggesting that the model is unable to effectively distinguish them based on the current feature set. One possible explanation is that each of those passes can shorten circuit depth or remove CNOTs, so the current features are not rich enough to differentiate them consistently. The \texttt{Miscellaneous} label performs the worst, reflecting its role as a catch-all class is not effective in this scenario, and alternative representations should be explored. Overall, the results show that unique structural signatures are learned well, whereas overlapping effects require richer features to achieve similar reliability.

\noindent \textbf{Benchmarking ML-models:} Table \ref{tab:benchmarks} shows the performance of various models on our 10,000-circuit dataset. The Hamming score represents the overlap between the predicted passes and the actual passes. It is calculated as the ratio of the intersection of predicted and actual passes (actual $\cap$ predicted) to the union of total passes (actual $\cup$ predicted) for a circuit sample. From the table, we see that the neural network achieves the highest Hamming score of approximately 0.68, meaning it correctly identifies 68\% of the passes. The Avg-F1 refers to the average F1-score across all classes, while the Micro-F1 calculates a global F1 score by summing the true positives, false positives, and false negatives across all classes. In both of these metrics as well, the neural network outperforms the other models, although there remains significant room for improvement.

\subsection{Discussions}

\noindent \textbf{Applications:} Besides reverse-engineering optimization techniques, the same trained classifier can be used as a lightweight compiler-forensics tool. By feeding it the (\texttt{original}, \texttt{compiled}) circuit pair as input, we can obtain a probabilistic fingerprint of which standard passes were likely executed. Users can then compare the predicted passes against the vendor’s stated optimization specifications to verify claims or detect unexpected transformations (e.g., a CNOT-cancellation stage that should have been disabled). Furthermore, since the model can flag unrecognized patterns (\texttt{Miscellaneous} label) with effective feature engineering, it can also help reveal proprietary or experimental optimizations, enabling deeper inspection without requiring source-level (i.e., white-box) access to the compiler itself. Thus, our work has the potential to bridge the gap between theoretical compilation algorithms and real-world implementations, offering developers crucial insights to optimize workflows.

\noindent \textbf{Limitations:} In this study, we test a subset of known optimization techniques to determine if a ``black-box'' compiler uses them in its circuit optimization algorithm. However, in practice, compilers may use completely novel or unknown techniques. Although we attempt to simulate such scenarios using our miscellaneous class, techniques falling entirely outside our predefined categories might still lead to mis-classification. Therefore, while our study effectively identifies known optimization techniques, it may not be suitable for detecting entirely novel methods.

\section{Conclusion}  
In recent years, there has been significant research evaluating the security of quantum computers. In this work, we focus on one such threat targeting quantum compilers. Due to the importance of compilers in overall circuit execution performance, they are vulnerable to confidentiality threats. Specifically, if an adversary can infer the inner workings, particularly the circuit optimization techniques of proprietary software, it could lead to critical issues, including IP theft. Here, we perform a preliminary study evaluating whether machine learning models can infer optimization techniques by analyzing only the original and optimized quantum circuits. Our findings indicate that adversaries can deduce some internal optimization techniques, highlighting the seriousness of this threat. Future works should focus on secure usage of quantum compilers to mitigate this risk.

\section*{Acknowledgment}
The work is supported in parts by NSF (CNS-2129675, CCF-2210963, OIA-2040667, DGE-1821766 and DGE-2113839), gifts from Intel and IBM Quantum Credits.

\bibliographystyle{ACM-Reference-Format}
\bibliography{refs}

\end{document}